\documentclass{jetpl}

\usepackage[T2A]{fontenc}
\usepackage[cp1251]{inputenc}
\usepackage[russian,english]{babel}

\twocolumn

\usepackage{color}

\lat

\title{Peculiarities of performance of the spin valve for the superconducting current}

\rtitle{Peculiarities of performance of the spin valve\ldots}

\sodtitle{Peculiarities of performance of the spin valve for the superconducting current}

\author{
P.\,V.~Leksin$^{\,+}$, A.\,A.~Kamashev$^{\,+}$, N.\,N.~Garif'yanov$^{\,+}$, I.\,A.~Garifullin$^{\,+}$\/\thanks{e-mail: ilgiz\_garifullin@yahoo.com},
Ya.\,V.~Fominov$^{\,*\nabla}$, J.~Schumann$^{\,\square}$, С.~Hess$^{\,\square}$, V.~Kataev$^{\,\square}$,
B.~B\"{u}chner$^{\,\square\times}$ }

\rauthor{P.\,V.~Leksin, A.\,A.~Kamashev, N.\,N.~Garif'yanov, et al.}

\sodauthor{Leksin, Kamashev, Garif'yanov, et al.}

\address{
$^+$Zavoisky Physical-Technical Institute, Russian Academy of Sciences, 420029 Kazan, Russia\\~\\
$^*$ L.\,D.~Landau Institute for Theoretical Physics RAS, 119334 Moscow, Russia\\~\\
$^\nabla$ Moscow Institute of Physics and Technology, 141700 Dolgoprudny, Russia\\~\\
$^\square$ Leibniz Institute for Solid State and Materials Research IFW Dresden, D-01171 Dresden, Germany\\~\\
$^\times$ Institut f\"{u}r Festk\"{o}rperphysik, Technische Universit\"{a}t Dresden, D-01062 Dresden, Germany
}

\dates{14 April 2013}{*}

\abstract{The spin valve effect for the superconducting current based on the superconductor/ferromagnet proximity effect has been studied for a
CoO$_x$/Fe1/Cu/Fe2/Cu/Pb multilayer. The magnitude of the effect $\Delta T_c =T_c^{\mathrm{AP}} -T_c^{\mathrm{P}}$, where $T_c^{\mathrm{P}}$ and $T_c^{\mathrm{AP}}$ are the
superconducting transition temperatures for the parallel (P) and antiparallel (AP) orientation of magnetizations, respectively, has been measured for
different thicknesses of the Fe1 layer $d_\mathrm{Fe1}$. The obtained dependence of the effect on $d_\mathrm{Fe1}$ reveals that $\Delta T_c$ can be
increased in comparison with the case of a half-infinite Fe1 layer considered by the previous theory. A maximum of the spin valve effect occurs at
$d_\mathrm{Fe1}\sim d_\mathrm{Fe2}$. At the optimal value of $d_\mathrm{Fe1}$, almost full switching from the normal to the
superconducting state when changing the mutual orientation of magnetizations of the iron layers Fe1 and Fe2 from P to AP is demonstrated.}


\begin{document}

\maketitle

The possibility to create a spin valve, based on the superconductor/ferromagnet (S/F) proximity effect is actively studied both theoretically and
experimentally. Two different constructions of the spin valve for the superconducting current have been theoretically proposed. The first
one \cite{Oh} is the F1/F2/S multilayer system where F1 and F2 are the ferromagnetic layers with uncoupled magnetizations, and S is the
superconducting layer. Calculations \cite{Oh} show that at parallel (P) orientation of magnetizations of F1 and F2 layers the superconducting transition
temperature $T_c$ ($T_c^\mathrm{P}$) is lower than in the case of their antiparallel (AP) orientation ($T_c^\mathrm{AP}$). The
second construction \cite{Tagirov,Buzdin99} is F1/S/F2. Its operation is similar to the first one. Several experimental works confirmed the
predicted effect of the mutual orientation of the magnetizations in the F/S/F structure on $T_c$ (see, e.g.,
\cite{Gu,You,Potenza,Pena,Moraru,Miao}).
However, the magnitude of the spin valve effect $\Delta T_c=T_c^\mathrm{AP}-T_c^\mathrm{P}$
turned out to be smaller than the width of the superconducting transition $\delta T_c$ itself. Hence a full switching between the normal and the
superconducting state was not achieved. Constructions  similar to that suggested in \cite{Oh} were studied to a less extent
\cite{Westerholt,Nowak,Nowak1}. Theoretical works by Fominov \textit{et al.} \cite{Fominov1,Fominov2} have generalized the theory of the
spin valve effect for both constructions taking into account the appearance of the triplet component in the superconducting condensate. Recently,
when studying the construction proposed by Oh \textit{et al.} \cite{Oh} on an example of multilayer CoO$_x$/Fe1/Cu/Fe2/In, we have
succeeded to obtain a full switching between the superconducting and the normal state when changing the mutual orientation of the magnetizations
of F1 and F2 layers \cite{Leksin1}. (Here CoO$_x$ is an antiferromagnetic bias layer which fixes the magnetization of the Fe1 layer along the
cooling field direction, Cu is a nonmagnetic layer N which decouples the magnetizations of the Fe1 and Fe2 layers, and In is a superconducting indium
layer). Furthermore, a detailed study of the spin valve effect has shown that the magnitude of the effect $\Delta T_c$ strongly
depends on the Fe2 thickness $d_\mathrm{Fe2}$ yielding the change of its sign at large values of $d_\mathrm{Fe2}$ \cite{Leksin2,Leksin3}.

To improve the operating parameters of the system (in particular, to increase $T_c$) we have replaced In by Pb \cite{Leksin4} and have found
out that the full switching can not be achieved  because of a large value of the superconducting transition width $\delta T_c$. It should be noted that the thickness of the
Fe1 layer should also affect the value of $\Delta T_c$. This is because the mean exchange field from two F-layers acting on Cooper pairs in the
space between Fe1 and Fe2 layers should be compensated for the AP orientation of the magnetizations of Fe1 and Fe2 layers. Thereby a naive
consideration shows that for the observation of the maximal spin valve effect it is desirable to have comparable values of $d_\mathrm{Fe1}$ and
$d_\mathrm{Fe2}$.

The basis of the present work has been formed by our earlier studies of the superconducting spin valve effect in the multilayer system
CoO$_x$/Fe1/Cu/Fe2/Cu/Pb \cite{Leksin4}. In the present paper we study a dependence of $\Delta T_c$ on the thicknesses of the Fe1 layer
$d_\mathrm{Fe1}$ and  of the Fe2 layer $d_\mathrm{Fe2}$. At the optimal value of $d_\mathrm{Fe1}$ we have succeeded to demonstrate
an almost full switching from the normal to the superconducting state.

The layer sequence CoO$_x$/Fe1/Cu/Fe2/Cu/Pb was deposited on a single-crystalline MgO substrate (Fig.~1).
\begin{figure}[h]
\centering{\includegraphics[width=0.6\columnwidth,angle=-90,clip]{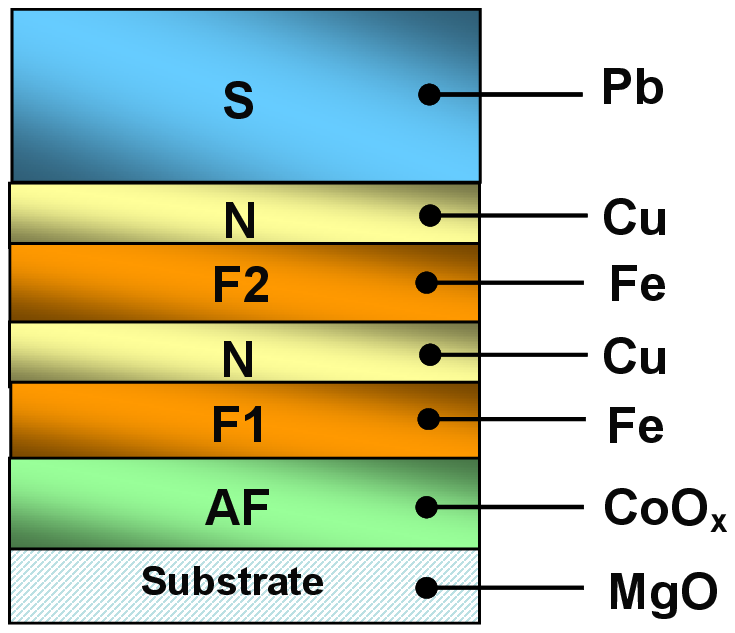}}
\caption{Fig.~1. Design of the studied samples.}
\end{figure}
We used the same sample preparation method, experimental setups, and protocols of magnetic and transport measurements as in our previous work
\cite{Leksin1}. An additional copper layer between Fe and Pb layers was evaporated in order to stabilize the properties of the Fe2/Pb interface.
Full characterization of these samples shows that this additional layer does not affect the spin valve effect. At the final step of the
preparation process all samples were capped by a 85\,nm thick SiN dielectric layer to protect the structure from oxidation.

For all samples we have performed magnetization measurements using a VSM SQUID (vibrating sample magnetometer superconducting quantum interference
device) magnetometer. We have measured the major and minor hysteresis loops $M(H)$ in order to determine the field range in which the full switching
between P and AP orientations of the magnetizations of the Fe1 and Fe2 layers is realized. We have found out that the hysteresis loop related to
the free iron layer Fe2 saturates at a field of the order of $\pm 1$\,kOe suggesting a complete suppression of the domain state.

Electrical resistivity measurements were performed with a standard four-point probe setup in the dc mode. We have combined the electrical setup
with a high homogeneous vector electromagnet that enables a continuous rotation of the magnetic field in the plane of the sample and have
used a system which enables a very accurate control of the real magnetic field acting on the sample. The magnetic field strength was measured
with an accuracy of $\pm 0.3$\,Oe using a Hall probe. The temperature of the sample was monitored by the 230\,$\Omega$ Allen-Bradley resister
thermometer which is particularly sensitive in the temperature range of interest. Therefore the accuracy of the temperature control within the same
measurement cycle below 2\,K was better than $\pm 2\div 3$\,mK. To avoid the occurrence of the unwanted out-of-plane component of the external field,
the sample plane position was always adjusted with an accuracy better than $3^{\circ}$ relative to the direction of the dc external field.

In order to study the influence of the mutual orientation of the magnetizations on $T_c$ we have cooled down the samples from room to a low
temperature at a magnetic field of 4\,kOe applied along the easy axis of the sample just as we did it when performing the SQUID magnetization
measurements. For this field both F-layers' magnetizations are aligned. Then at the in-plane magnetic field value of $\pm H_0 = \pm 1$\,kOe the
temperature dependence of the resistivity $R$ was recorded.

Fig.~2 depicts the dependence of the magnitude of the spin valve
effect $\Delta T_c=T_c^\mathrm{AP}-T_c^\mathrm{P}$ on the
thickness of the Fe2 layer for fixed thickness of the Fe1 layer $d_\mathrm{Fe1}=2.5$\,nm \cite{ref}.
\begin{figure}[h]
\centering{\includegraphics[width=0.6\columnwidth,angle=-90,clip]{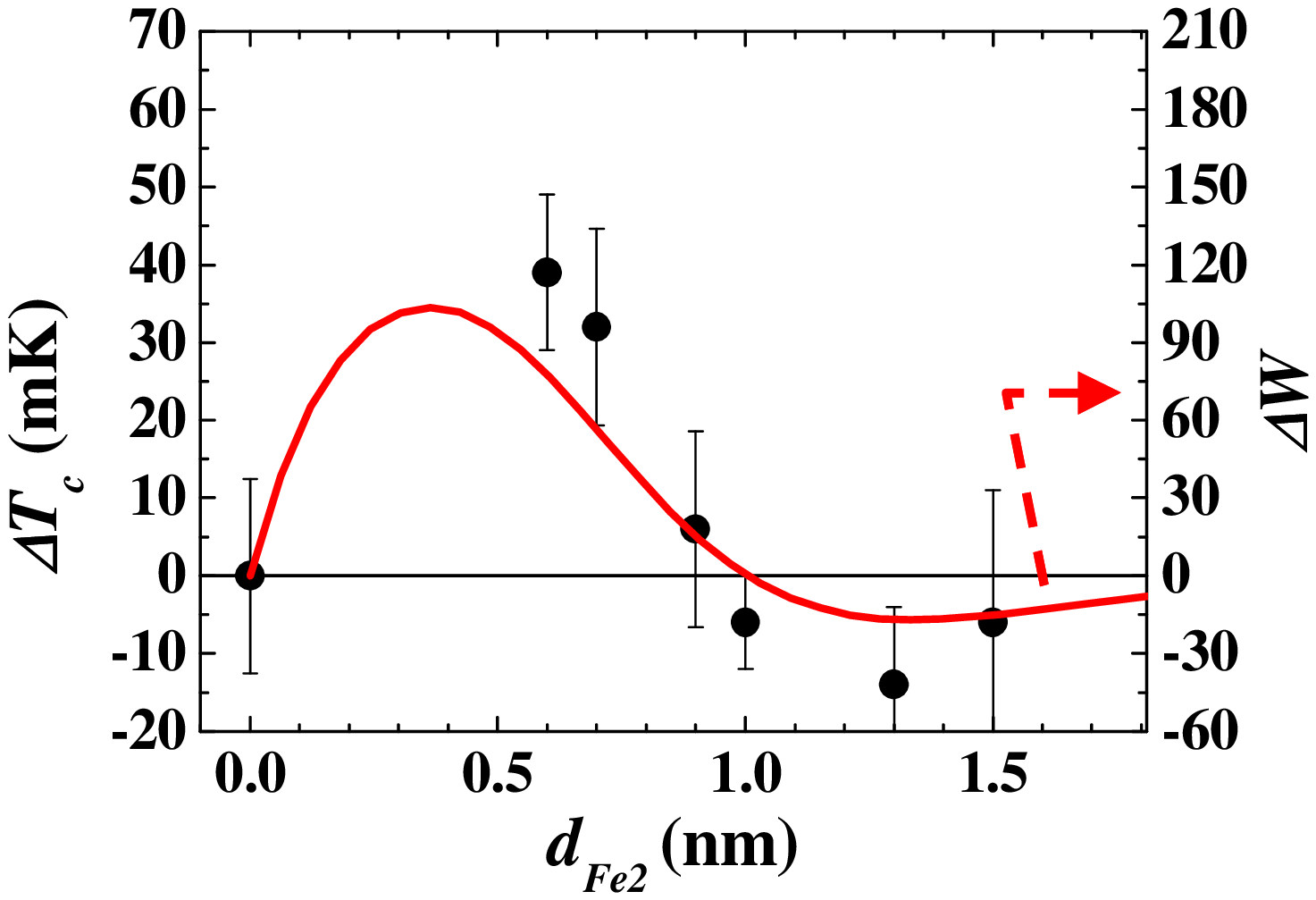}}
\caption{Fig.~2. Dependence of the magnitude of the spin
valve effect $\Delta T_c$ on the thickness of the Fe2 layer at a
fixed values of the S layer $d_\mathrm{Pb}=35$\,nm and Fe1 layer $d_\mathrm{Fe1}=2.5$\,nm.  Solid line is a
theoretical curve for $\Delta W$ (see the text).}
\end{figure}
For samples with $d_\mathrm{Fe2}<0.95$\,nm we have observed the direct effect with
$T_c^\mathrm{P}<T_c^\mathrm{AP}$, whereas for samples with $d_\mathrm{Fe2}>0.95$\,nm the inverse effect with
$T_c^\mathrm{P}>T_c^\mathrm{AP}$ has been found.
 The data shown in Fig.~2 are qualitatively rather
similar to our previous results on the sign changing oscillating spin valve effect in CoO/Fe1/Cu/Fe2/In multilayers \cite{Leksin1,Leksin2,Leksin3}.
Practically, in our samples the iron layers thinner than 0.5\,nm  are not continuous anymore. The reason why $\delta T_c$ for the Pb layer
in contact with the Fe layer is larger than that for the In layer in contact with the same Fe layer is unclear.

Fig.~3 shows the dependence of the magnitude of the spin valve
effect $\Delta T_c=T_c^\mathrm{AP}-T_c^\mathrm{P}$ on the
thickness of the Fe1 layer for two fixed thicknesses of the Fe2 layer.
\begin{figure}[h]
\centering{\includegraphics[width=0.6\columnwidth,angle=-90,clip]{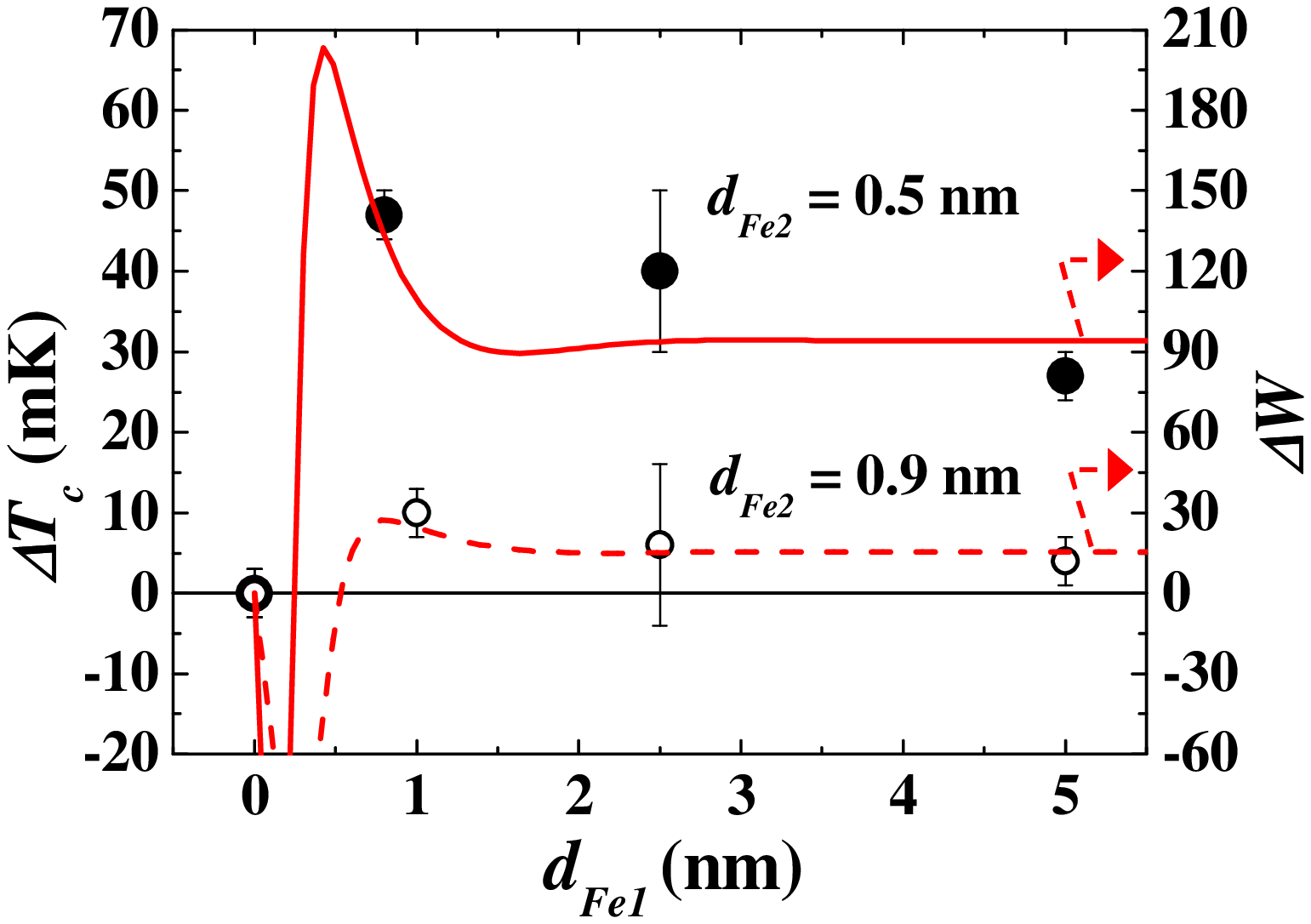}} \caption{Fig.~3. The dependence of the $T_c$ shift $\Delta
T_c=T_c^\mathrm{AP}-T_c^\mathrm{P}$ on the Fe1 layer thickness $d_\mathrm{Fe1}$ for the series of the samples with $d_\mathrm{Fe2}=0.5$\,nm ($\bullet$) and
0.9\,nm ($\circ$) at fixed $d_\mathrm{Pb}=35$\,nm. The applied switching field $H_0=\pm 1$ kOe lies in the plane of the film. Solid and dashed lines are theoretical
curves for $\Delta W$ (see the text).}
\end{figure}
One can see that the dependences of $\Delta T_c (d_\mathrm{Fe1})$ have maximum at the values of $d_\mathrm{Fe1}$ of the order of 1 nm or less.
The maximum in $\Delta T_c (d_\mathrm{Fe1})$ can be related to the compensation of the mean
exchange field in the space between Fe1 and Fe2 layers occurring for nearly equal thicknesses of these two layers. With increasing the
$d_\mathrm{Fe2}$ value the spin valve effect diminishes. This is because the penetration depth of the Cooper pairs into the iron layer is $\xi_F
\simeq0.8$\,nm \cite{Lazar}. That means that only a small amount of the Cooper pairs can penetrate through the Fe2 layer to be subjected to the
influence of the Fe1 layer.

For the spin valve sample CoO$_x$/Fe1(0.8 nm)/Cu(4 nm)/Fe2(0.5 nm)/Cu(1.2 nm)/Pb(60 nm) the
difference in $T_c$ for different magnetic field directions is
clearly seen (see Fig.~4).
\begin{figure}[h]
\centering{\includegraphics[width=0.6\columnwidth,angle=-90,clip]{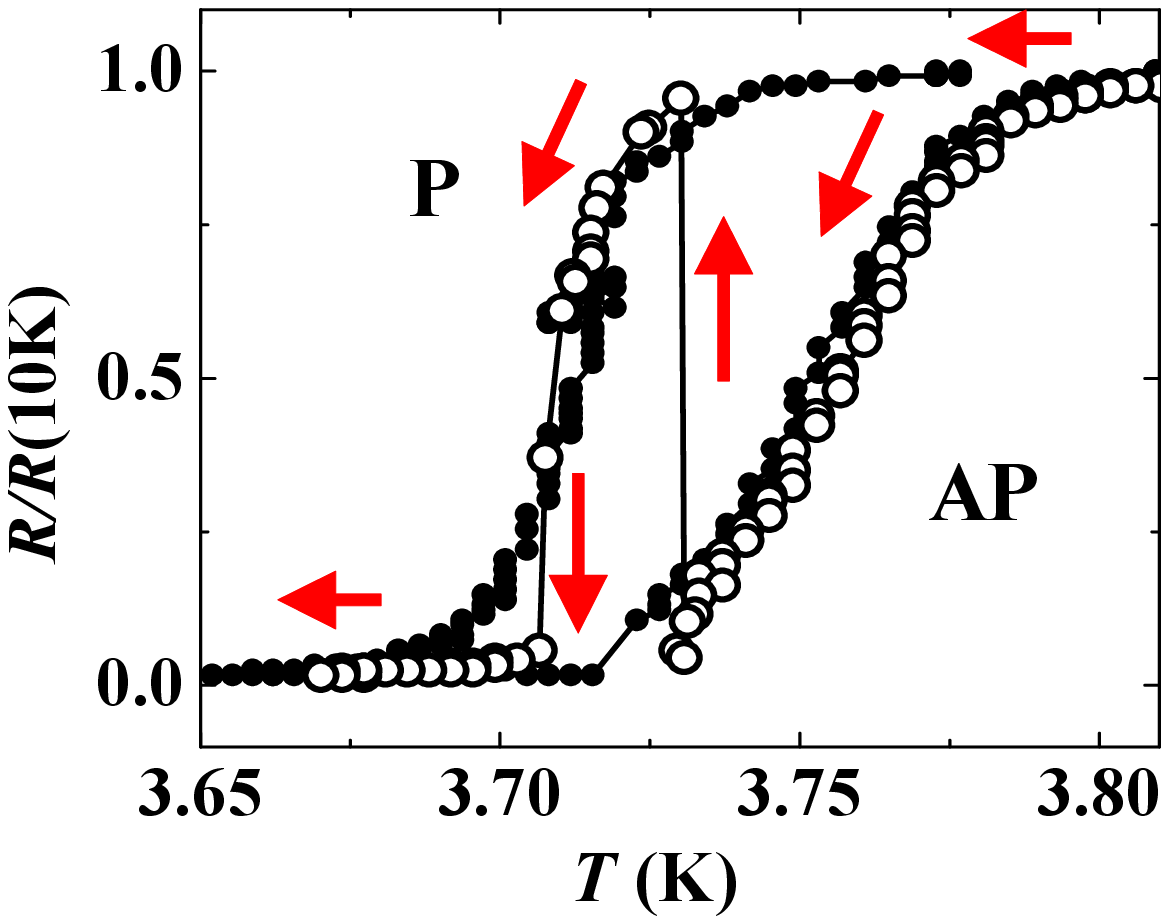}} \caption{Fig.~4. ($\bullet$) - Superconducting transition
 curves for P ($H_0=+1$\,kOe) and AP ($H_0=-1$\,kOe) orientations of the Fe1 and Fe2 layers' magnetizations, respectively, for the sample CoO$_x$/Fe1(0.8 nm)/Cu(4 nm)/Fe2(0.5 nm)/Cu(1.2 nm)/Pb(60 nm).
 ($\circ$) - Instant switching between superconducting state and normal state by switching between AP ($H_0=-1$\,kOe) and P ($H_0=+1$\,kOe) orientations of the Fe1 and Fe2 layers' magnetizations during a slow temperature sweep.}
\end{figure}
The superconducting transition temperature for the AP orientation of the magnetizations occurs at a temperature exceeding $T_c$ for the P
orientation of the sample by 40\,mK, which is of the order of the superconducting transition width $\delta T_c$. This opens a possibility to
switch off and on the superconducting current flowing through our samples \emph{almost completely} within the temperature range corresponding to the
$T_c$-shift by changing the mutual orientation of magnetization of F1 and F2 layers. To demonstrate this we have performed resistivity
measurements of the sample by sweeping slowly the temperature within the $\Delta T_c$ and switching the magnetic field between $+1$\, kOe
and $-1$\,kOe (Fig.~4).


Let us discuss the results shown in Figs.~2 and~3 in the framework of the S/F proximity effect theory. For that we extend the results of
\cite{Fominov2} where $T_c$ of an F1/F2/S trilayer was considered in the simplest formulation: all the interfaces were assumed to be
transparent, and the outer F layer was assumed to be half-infinite. That formulation enabled studying interplay of different magnetizations'
orientations in the two F layers. For the present experiment, we are interested in theoretical results only for the two collinear orientations
(parallel and antiparallel). At the same time, we should extend the previous theory \cite{Fominov2} for arbitrary thickness of the outer F layer (Fe1
layer in notation of the present paper).

In this theory, the set of equations describing superconductivity in all parts of the structure is reduced to the problem for the S layer only. All
the information about the rest of the structure (two F layers and interfaces) is then contained in a real-valued parameter $W$ that enters the
effective boundary condition for the F2/S interface: $\xi (df_0/dx)=Wf_0$, where $f_0$ is the singlet component of the anomalous Green function in
the S layer and $\xi$ is the coherence length. Physically, $W$ determines how strongly superconductivity in the S layer is suppressed by the rest of
the structure due to the proximity effect. The larger $W$ is, the stronger $T_c$ is suppressed. Therefore, $\Delta T_c$ should correlate with $\Delta
W = W^\mathrm{P}-W^\mathrm{AP}$. Note at the same time that generally there is no simple explicit
relation (like, e.g., proportionality) between them.

In the parallel configuration, we effectively have F/S system with a single F layer of thickness $d_\mathrm{Fe1}+d_\mathrm{Fe2}$. Then we reproduce the result of \cite{FCG}:
\begin{multline} \label{W0}
W^\mathrm{P} = 2 k_h \xi \frac{\sigma_F}{\sigma_S} \times \\
\times \frac{\cosh(d_1+d_2) - \cos(d_1+d_2)}{\sinh(d_1+d_2) - \sin(d_1+d_2)+2 \kappa \tanh(k_\omega d_S)},
\end{multline}
where
\begin{equation}
d_1 = 2k_h d_\mathrm{Fe1},\quad d_2 = 2 k_h d_\mathrm{Fe2},\quad \kappa = \frac{\sigma_S k_\omega}{\sigma_F k_h}.
\end{equation}
Here $\sigma_S$ and $\sigma_F$ are the normal-state conductivities of the S and F layers, respectively (the diffusion constants will be denoted $D_S$
and $D_F$); $\xi= \sqrt{D_S/2\pi T_{cS}}$ is the coherence length for the S layer (where $T_{cS}$ is the critical temperature in the bulk). The wave
vectors $k_\omega = \sqrt{2\omega/D_S}$ and $k_h = \sqrt{h/D_F}$ describe the scales of the spatial inhomogeneity due to the proximity effect in the
S and F layers, respectively ($\omega$ is the Matsubara frequency, which can be taken for estimates as $\pi T_c$). We assume that $T_c/h$ is small
enough so that $\kappa \ll 1$.

In the antiparallel configuration, we generalize treatment of \cite{Fominov2} taking into account finite $d_\mathrm{Fe1}$ (and also taking into
account different degree of disorder in the S and F layers, so that $D_S$, $\sigma_S$ and $D_F$, $\sigma_F$ are different). The result is
\begin{equation} \label{WPi}
W^\mathrm{AP} = 2 k_h \xi \frac{\sigma_F}{\sigma_S} \frac{\mathcal N}{\mathcal D},
\end{equation}
where the numerator and denominator of the last fraction are
\begin{align}
\mathcal N &= \cosh d_1 \cosh d_2 - \cos d_1 \cos d_2 - \notag \\
&- \sin d_1 \sinh d_2 - \sinh d_1 \sin d_2, \\
\mathcal D &= \cosh d_1 \sinh d_2 + \sinh d_1 \cos d_2 - \notag \\
& - \sin d_1 \cosh d_2 - \cos d_1 \sin d_2 + 2 \kappa \tanh(k_\omega d_S).
\end{align}


The qualitative correlation between measured $\Delta T_c$ and calculated $\Delta W$ in Figs.~2 and~3 looks satisfactory. Note that $\Delta
T_c(d_\mathrm{Fe2})$ and $\Delta W(d_\mathrm{Fe2})$ in Fig.~2 approximately correspond to the limit of infinite $d_\mathrm{Fe1}$ since this thickness
is several times larger than the penetration depth $k_h^{-1} \equiv \xi_F$. The central result of our paper, the dependence
of $\Delta T_c$ on $d_\mathrm{Fe1}$, is shown in Fig.~3. For the calculated curves of $\Delta W$ we used the same parameters as in our previous paper
\cite{Leksin4}: for the F layers we put the Fermi velocity $v_F = 2\cdot 10^8$\,cm/s, mean free path of conduction electrons $l_f = 1.5$\,nm, and the
exchange field $h = 0.85$\,eV. According to our calculations $\kappa = 0.009$. The dirty-limit theory (the Usadel equations) that we used to calculate $\Delta W$, requires, in particular, the
condition $h l_f /\hbar v_F \ll 1$ (where $\hbar$ is Planck's constant) \cite{comment}. The above-mentioned parameters correspond to the value of $h
l_f /\hbar v_F$ of the order of one. There are also other simplifications of the theory compared to the experiment; the main of them is probably that
all interfaces are assumed to be fully transparent [while the exchange splitting of the conduction band of the ferromagnet is at least one source for
a non-perfect F/S interface transparency (see, e.g., \cite{Lazar})]. Taking this into account, we can only expect qualitative agreement between
theory and experiment.

Let us now discuss our qualitative understanding of the spin-valve effect dependence on $d_\mathrm{Fe1}$, keeping in mind that $k_h^{-1}$ is the
penetration depth of superconducting correlations in the ferromagnet. In the parallel orientation, the two F layers act as a singe layer leading to
some suppression of superconductivity in the S part. The spin-valve effect is due to partial mutual compensation of the exchange fields of the two F
layers when they are in the antiparallel orientation. We expect that the spin-valve effect is largest when this compensation is most effective, which
is the situation at $d_\mathrm{Fe1} \sim d_\mathrm{Fe2}$ if $d_\mathrm{Fe2} < k_h^{-1}$. At $d_\mathrm{Fe2}> k_h^{-1}$ this condition should be
modified since the outer layer of the same thickness, $d_\mathrm{Fe1} > k_h^{-1}$ cannot fully participate in the compensation effect. Only a part of
it, with thickness of the order of $k_h^{-1}$, is effective, so we expect the spin-valve effect in this case to be strongest for $d_\mathrm{Fe1} \sim
k_h^{-1}$. The theoretical results for $\Delta W$ presented above confirm this qualitative picture, and the experimental data in Fig.~3,
corresponding to $k_h^{-1} \approx 0.8$\,nm, seem to be consistent with it. Note that the theory also predicts peculiarities of the spin-valve effect
at very small $d_\mathrm{Fe1}$ (due to interference features of the oscillating proximity effect in the F part), however we do not focus on them
since we do not have experimental data for such small thicknesses.

In conclusion, we employed a spin valve system  CoO$_x$/Fe1/Cu/Fe2/Cu/Pb to investigate the dependence of the magnitude of the spin valve effect
$\Delta T_c$ on the thickness of the Fe1 layer $d_\mathrm{Fe1}$. We have observed that the $\Delta T_c$ value can be slightly (by $\sim 20 \%$)
increased in comparison with the case of the half-infinite Fe1 layer considered by the theory \cite{Fominov2}. The achieved theoretical
understanding of the $\Delta T_c(d_{\mathrm{Fe1}})$ dependence warrants a next step towards practical application. Indeed, the optimal choice of the
thicknesses $d_{\mathrm{Fe1}}$ and of $d_{\mathrm{Fe2}}$ gives a possibility to get an almost full switching from the normal to the superconducting state.

The work was partially supported by the DFG (grant BU 887/13-2) and by the RFBR (grants Nos.\ 11-02-01063-a and 13-02-97037-r-povolzhje\_a). Ya.V.F.\
was supported by the RFBR (grant No.\ 11-02-00077-a), the Ministry of Education and Science of Russia (Contract No.\ 8678), and the program ``Quantum
mesoscopic and disordered structures'' of the RAS.


\begin{thebibliography}{99}
\bibitem{ref}
Experimental points on this figure are taken from our previous paper \cite{Leksin4}. However, the theoretical curve is slightly changed in order to
describe simultaneously the experimental results shown in Figs.~2 and~3.
\bibitem{comment}
Note that although the Usadel equations are formally not valid in the case of strong ferromagnets, it turns out that often they can describe experimental results surprisingly well (see, e.g., \cite{Moraru}).
\end{thebibliography}
\end{document}